\documentclass[final,5p,twocolumn]{elsarticle}
\usepackage{amssymb}
\usepackage{graphicx}
\usepackage{geometry}
\usepackage[dvipsnames,table]{xcolor}
\usepackage{slashed}
\usepackage{hyperref} 
\usepackage{xspace}
\usepackage[tight]{subfigure}
\usepackage{amsmath}
\usepackage{amsfonts}
\usepackage{mathrsfs}
\usepackage{comment}
\usepackage{afterpage}
\usepackage{verbatim}
\usepackage{booktabs}
\usepackage{array}
\usepackage[title,titletoc]{appendix}
\usepackage{tabularx}
\newcolumntype{C}[1]{>{\centering\arraybackslash}p{#1}}

\newcommand{\beq}{\begin{eqnarray}}
\newcommand{\eeq}{\end{eqnarray}}
\newcommand{\nnb}{\nonumber}

\newcommand{\GeV}{{\,\rm GeV}}
\newcommand{\wwcom}{$WW$ CoM}
\newcommand{\wzcom}{$WZ$ CoM}

\journal{Physics Letters B}

\begin{document}

\begin{frontmatter}
  
  \title{
   Different polarization definitions in same-sign $WW$ scattering at the LHC.
  }
  \author[label1]{Alessandro Ballestrero}\ead{ballestr@to.infn.it}
  \author[label1,label2]{Ezio Maina}\ead{maina@to.infn.it}
  \author[label3]{Giovanni Pelliccioli}\ead{giovanni.pelliccioli@physik.uni-wuerzburg.de}
  \address[label1]{INFN, Sezione di Torino, via Pietro Giuria 1, 10125 Torino (Italy)}
  \address[label2]{University of Torino, Department of Physics, via Pietro Giuria 1, 10125 Torino (Italy)}
  \address[label3]{University of W\"urzburg, Instit\"ut f\"ur Theoretische Physik und Astrophysik,
  Emil-Hilb-Weg 22, 97074 W\"urzburg (Germany)}

\begin{abstract}
We study the polarization of positively charged $W$'s in the scattering of massive electroweak bosons at 
hadron colliders. 
We rely on the separation of weak boson polarizations in the gauge-invariant, doubly-resonant part of the 
amplitude in Monte Carlo simulations. Polarizations depend on the reference frame in which they are defined.
We discuss the change in polarization fractions and in kinematic distributions
arising from defining polarization vectors in two different 
reference frames which have been employed in recent experimental analyses.
\end{abstract}
\begin{keyword}
  Vector Boson Scattering \sep LHC \sep Polarization \sep Electroweak %  \PACS VBSCAN-PUB-05-20
\end{keyword}
\end{frontmatter}

\section{Introduction}\label{intro}
Vector Boson Scattering  (VBS) at the LHC represents
a crucial process for both Standard Model (SM) analyses and for Beyond-the-Standard-Model
(BSM) searches. The $W^\pm W^\pm$ channel with leptonic decays features the largest
signal-to-background ratio among VBS processes.
It has been measured \cite{Aad:2014zda,Sirunyan:2017ret,Aaboud:2019nmv,Sirunyan:2020gvn} and studied
\cite{Khachatryan:2014sta,Aaboud:2016uuk,Sirunyan:2019der,Sirunyan:2020gyx} in LHC
proton-proton collisions with $\sqrt{s} =$ 8 and 13 TeV,
and the luminosity increase expected for the next runs will yield
more accurate measurements \cite{CMS-PAS-FTR-18-014,CMS:2018mbt,Azzi:2019yne}.

A detailed study of  VBS  with two positively-charged leptons and missing transverse
momentum in the final state is presented in Ref.~\cite{Ballestrero:2018anz}. 
Predictions include NLO QCD corrections and are matched to parton showers.

The importance of VBS with longitudinal bosons is related to the delicate cancellation of large,
unitarity-violating contributions
in the SM, that come respectively from pure gauge and from Higgs diagrams. 
Any modification of the SM realization of the Electroweak Symmetry Breaking
mechanism (EWSB) could interfere with this delicate cancellations in VBS. Therefore, it
is essential to devise a good definition of polarized signals at the theoretical level,
and to investigate their phenomenology, in order to identify observables which allow
the efficient experimental separation of polarized processes at the LHC.

Both CMS  and ATLAS have measured the $W$ polarization fractions 
in the $W+\,$jets \cite{Chatrchyan:2011ig,ATLAS:2012au} channel and in $t\,\bar{t}$ events 
\cite{Aaboud:2016hsq, Khachatryan:2016fky,Aad:2020jvx}.
$Z$ polarization fractions at the LHC have been measured in \cite{Aad:2016izn,Khachatryan:2015paa}.
The first
polarization measurement at 13 TeV has been performed by
ATLAS in $WZ$~production \cite{Aaboud:2019gxl}.

In Refs.~\cite{Ballestrero:2017bxn,Ballestrero:2019qoy} we extensively investigated the $W^+W^-$, $W^+Z$ 
and $Z\!Z$ scattering channels.
In the fully-leptonic $W^+W^+$ scattering, like in $W^+W^-$,
the presence of two neutrinos in the final state inhibits the experimental
reconstruction of the center-of-mass frame of each $W$ boson.
This strongly limits the search for kinematic variables which are sensitive to vector boson polarizations.
A more in-depth understanding of the polarization structure of $W^+W^+$ is urgent.
Although a number of results are reported in Ref.~\cite{Pelliccioli:2019wlv, JanssenVBSCAN},
a detailed study of same sign $W$'s
scattering, in the spirit of Refs.~\cite{Ballestrero:2017bxn,Ballestrero:2019qoy}, is still missing.

The predictions presented here are not fully
realistic, since we are considering only the leading order SM electroweak signal for VBS,
without including the QCD background, higher-order corrections and parton-shower effects.
This paper could serve as a benchmark study in view of more precise investigations.

Two are the novel features with respect to Refs.~\cite{Ballestrero:2017bxn,Ballestrero:2019qoy}. 
First, we focus on doubly-polarized distributions;
second, we investigate the effect of defining the polarization vectors in different reference frames.
We are further motivated by the fact that the experimental analyses in Refs. 
\cite{Chatrchyan:2011ig,ATLAS:2012au,Aaboud:2016hsq, 
Khachatryan:2016fky,Aad:2020jvx,Aad:2016izn,Khachatryan:2015paa,Aaboud:2019gxl} employ
distinct definitions of polarizations.

The paper is organized as follows. In Sect.~\ref{theo} we recall quickly the theoretical
concepts which are needed to separate polarization modes of weak bosons
in a VBS context. After describing the setup of our simulations in Sect.~\ref{setup}, and
showing a few results in the absence of lepton cuts in Sect.~\ref{inclusive}, we present
integrated cross-sections and relevant distributions for polarized $W$ bosons in the presence
of realistic lepton cuts (Sect.~\ref{leptoncut}). We focus on observables with discriminating
power among different polarization modes, and on the shift in polarization fractions arising from defining
polarization vectors in two different reference frames. In Sect.~\ref{otherproc} we briefly address the effect
due to different polarization vector definitions in other VBS channels ($W^+W^-,\,W^+Z$ and
$ZZ$), limiting ourselves to integrated cross-sections. Finally, in Sec.~\ref{concl} we draw our conclusions.

\section{Defining polarized signals at the LHC}\label{theo}
Electroweak massive bosons feature three physical polarization modes, longitudinal, left- and right-handed.
The production of a boson with definite polarization state is well-defined only for an on-shell boson. 
However, being unstable particles, $W$'s and $Z$'s decay into leptons or quarks before detection.
A possible solution is to split the numerator of a $W/Z$ boson propagator
into a sum of terms, one for each physical polarization state: this statement holds for a general $\xi$-gauge 
choice, if
weak boson decay products are massless. Selecting a single term in the sum,
corresponds to selecting a definite polarization state of the propagating boson.

An additional complication is the non-resonant character of many diagrams
in multi-boson processes already at tree-level. For
$W^+W^+$ scattering at the LHC, doubly-resonant, singly-resonant and non-resonant
diagrams contribute at $\mathcal{ O} (\alpha^6)$. The latter two classes of diagrams don't
expose a factorized structure (production $\times$ propagator $\times$ decay), therefore
for them separating polarized terms is simply impossible.
An approximate  solution to this problem is to consider doubly-resonant diagrams only, and treating
them with a double-pole approximation \cite{Denner:2000bj} to recover $SU(2)_L\times U(1)_Y$ gauge invariance.
More specifically we employ the same on-shell projection technique {(OSP)} introduced for $W^+W^-$
in Ref.~\cite{Ballestrero:2017bxn}.

For the aim of this paper, it is essential to recall that the definition of polarizations
is not unique, since polarization vectors are not  Lorentz
covariant. For a boson with momentum $p$ and polarization state $\lambda$, and a generic
Lorentz transformation $\Lambda$, it can be proved
that $\varepsilon^\mu_\lambda(\Lambda\cdot p) \neq {\Lambda^\mu}_\nu\, \varepsilon^\nu_\lambda(p)$. 
Therefore, we
need to specify the reference frame in which polarization vectors are computed.
In Refs.~\cite{Ballestrero:2017bxn,Ballestrero:2019qoy},
they are defined in the laboratory, which is a natural choice at the LHC.
Recently, the ATLAS Collaboration measured
polarizations in $WZ$-pair production \cite{Aaboud:2019gxl}, defining
polarization observables in the boson-boson center-of-mass frame. 
This choice has the advantage that 
the lines of flight of the two bosons, whose direction determines the longitudinal polarization vectors,
are equal, up to a change of sign.
It is also more directly related to the weak boson scattering process embedded in the hadronic process and
therefore, it might be better suited to the search for new physics affecting the EWSB mechanism.
Phenomenological studies
with polarization vectors defined in the boson-boson center-of-mass have recently appeared
\cite{Baglio:2019nmc}.

In the following we discuss singly- and doubly-polarized integrated cross-sections and distributions, 
comparing the results with polarization vectors defined in:\\[-0.6cm]
\begin{itemize}
\item the laboratory frame (Lab);\\[-0.6cm]
\item the center-of-mass frame of the two $W^+$ bosons (\wwcom),\\[-0.6cm]
\end{itemize}
focusing on how the two different
definitions affect the distributions of kinematic variables that are observable at the LHC.

The possibility of defining the polarization vectors in the Lab or in the \wwcom~has been recently introduced
in \texttt{PHANTOM} \cite{Ballestrero:2007xq}. Different reference frames 
to define polarizations are available in \texttt{MADGRAPH5} \cite{BuarqueFranzosi:2019boy}.

\section{Setup}\label{setup}
The process under investigation is $p p \rightarrow j j e^+ {\nu_e} \mu^+ \nu_\mu$
with center-of-mass energy of 13 TeV. All the
total cross-sections and distributions shown below have been computed at parton-level 
with the \texttt{PHANTOM} \cite{Ballestrero:2007xq} Monte Carlo, using
\texttt{NNPDF3.0} PDFs \cite{Ball:2014uwa} computed at LO. The
factorization scale is set to $\mu_{\rm F} = {({p_t^{j_1}\,p_t^{j_2}})}^{1/2}$. We only consider
leading-order pure electroweak contributions, $\mathcal{O}(\alpha^6)$, which
are usually considered as the VBS signal, in contrast with the QCD background
($\mathcal{O}(\alpha^4\alpha_{\rm s}^2)$). We employ the Complex-Mass Scheme
\cite{Denner:1999gp,Denner:2005fg} for SM couplings and masses in full computations,
while {the OSP} results rely on real couplings with vanishing weak boson widths (apart from
the widths appearing in the off-shell propagators denominators corresponding to projected
$W^+$'s). We use the following pole masses and widths for weak bosons:
$M_{W}=80.358\GeV$, $M_{Z}=91.153\GeV$, $\Gamma_{W}=2.084\GeV$, $\Gamma_{Z}=2.494\GeV$.
The Fermi constant is set to $G_\mu=1.16637 \cdot 10^{-5}\GeV^{-2}$.

The following selection cuts are understood for all results:\\[-0.6cm]
\begin{itemize}
\item maximum jet pseudorapidity, $|\eta_j| < 5$;\\[-0.6cm]
\item minimum jet transverse momentum, $ p_t^j > 20$ GeV;\\[-0.6cm]
\item minimum jet--jet invariant mass, $M_{jj} > 500$ GeV;\\[-0.6cm]
\item minimum jet--jet pseudorapidity separation, $|\Delta\eta_{jj}| > 2.5$.\\[-0.6cm]
\end{itemize}
These selections define the inclusive setup, which is considered in Sect.~\ref{inclusive}.
In Sect.~\ref{leptoncut}, three additional
cuts are applied to the simulated events:\\[-0.6cm]
\begin{itemize}
\item maximum lepton pseudorapidity, $|\eta_{\ell}| < 2.5$;\\[-0.6cm]
\item minimum lepton transverse momentum, $ p_t^{\ell} > 20$ GeV;\\[-0.6cm]
\item minimum missing transverse momentum, $p_t^{\rm miss} > 40$ GeV.\\[-0.6cm]
\end{itemize}
Applying OSP requires
the mass of the four lepton system to be larger than twice the $W$
pole-mass, otherwise the technique is not defined. It is worth noting
that the number of events below this threshold is roughly 0.6\% in the
full calculation. This means that comparing full results computed over the whole range of
$M_{4\ell}$ and OSP results (polarized and unpolarized) computed for
$M_{4\ell}> 161\GeV$ does not introduce any relevant bias. This is
the procedure adopted in the following.

\section{Validation in the absence of lepton cuts}\label{inclusive}
In this section we concentrate on the inclusive setup, as defined above.
These results, without selection cuts on leptons, are not realistic, but useful to understand the quality of the
signal definition and to get a first feeling of the polarization structure in the scattering process.

In Tab.~\ref{table:sigmaincl} we show the total cross-sections for the unpolarized, singly-polarized and doubly-polarized signal,
with the two polarization definitions (in the Lab and in the \wwcom).
\begin{table}[th]
\begin{center}
\begin{tabular}{|C{1.4cm}||C{1.65cm}|C{1.65cm}|C{1.2cm}|}
\hline
\cellcolor{ForestGreen!9}   & \cellcolor{ForestGreen!9}  Lab  & \cellcolor{ForestGreen!9} \wwcom &  \cellcolor{ForestGreen!9} ratio    \\
\hline
\hline
full   & \multicolumn{2}{c|}{3.185(3)} & -  \\
\hline                                  
unpol & \multicolumn{2}{c|}{3.167(2)} & -  \\
\hline                                  
\hline                                 
0-unpol & 0.8772(8) & 0.8374(9)        &  0.95 \\
\hline                                  
T-unpol & 2.287(2) & 2.329(2)          &  1.02 \\
\hline                                  
\hline                                  
0-0 &  0.2573(3) & 0.3275(4)            &  1.27 \\
\hline                                  
0-T,T-0&  0.6199(6) & 0.5081(5)            &  0.82 \\
\hline                                 
T-T & 1.666(1) & 1.820(1)               &  1.09 \\
\hline
\end{tabular}
\end{center}
\caption{Total cross-sections (fb) for $W^+W^+$ scattering in the absence of lepton cuts.
  In the first column ``0'' stands for longitudinal, ``T'' for transverse, ``unpol'' for
  unpolarized. The first label is relative to the $W^+$ that decays into $e^+\nu_e$, the
  second one to the $W^+$ decaying into $\mu^+\nu_\mu$. The ratios are computed as polarized
  results in the \wwcom~ over those in the Lab. Number in parentheses are numerical MC errors.}
\label{table:sigmaincl}
\end{table}
As a first comment the unpolarized OSP result reproduces at the sub-percent level the full
cross-section. The good behavior of the OSP approximation is further confirmed at the differential
level in most of the distributions we have considered.

The singly-polarized cross-sections are similar to each other in the two definitions. The
longitudinal one decreases by 5\% passing from the Lab to the \wwcom, while
the transverse one increases by 2\%. Note that the sum of longitudinal and transverse polarizations
approximates the unpolarized result to better than 1\% with both definitions.
The longitudinal fraction is 28\% of the total in the Lab, 26\% in the \wwcom.

The differences between the polarization fractions computed in the two reference frames are larger for
the doubly-polarized results. In the \wwcom, the longitudinal-longitudinal and transverse-transverse
cross-sections are 27\% and 9\% larger, respectively,  with respect to the corresponding
ones in the Lab, while the mixed contributions are 18\% smaller.
Therefore, the \wwcom~ definiton might be more useful in extracting the
longitudinal-longitudinal contribution from the unpolarized process.

Given the large number of final state particles, one could na\"ively expect that the doubly-polarized fractions
could be obtained as
products of singly-polarized ones. This would be true if the two boson spin states were
not correlated. In the Lab, this would give a longitudinal-longitudinal cross-section of 0.24 fb which
is not so far from the Monte Carlo result (0.26 fb). The mixed contributions would both be 0.64 fb,
and the doubly-transverse 1.67 fb, to be compared with the actual values 0.62 fb and 1.67 fb, respectively.
In the Lab , the correlation
between the two bosons polarizations is mild, and its effect is hardly visible in configurations which involve 
transverse bosons.

The correlation is much more evident in the \wwcom. 
Under the zero-correlation hypothesis, the doubly-longitudinal
rate would amount to 0.22 fb, which is
35\% lower than the value computed by the Monte Carlo.

We have tested the validity of the polarized signals comparing Monte Carlo results with
those extracted analytically from the angular distributions of the decay leptons.
Using the analytic expression for the $W^+$ LO differential decay rate in the lepton angle $\theta^*_{\ell}$ (computed in the $W$ rest frame),
\beq
&&\hspace{-1.2cm} \frac{1}{\sigma}\frac{d\sigma}{d\cos\theta^*_{\ell^+}}\,=\,
  \frac{3}{4}f_0\sin^2\theta^*_{\ell^+} + \frac{3}{8}f_L(1-\cos\theta^*_{\ell^+})^2 +\nnb\\
 &&\hspace{0.8cm} +\frac{3}{8}f_R(1+\cos\theta^*_{\ell^+})^2\,,
\eeq
one can easily extract the polarization fractions $f_0,f_L,f_R$
projecting the unpolarized $\cos\theta^*_{\ell^+}$ distribution
onto the first three Legendre polynomials \cite{Ballestrero:2017bxn}.
If the polarizations are defined in the Lab (\wwcom), $\theta^*_{\ell^+}$ must be computed
with respect to the $W^+$ direction in the  Lab (\wwcom) frame.

Note that this method can be applied to extract single polarization fractions from the full distribution
as well as to extract double polarization fractions from singly-polarized distributions.
We have performed all these checks, for both polarization definitions, finding perfect agreement with the 
distributions  computed directly with the Monte Carlo, both in normalization and in shape.
As a last comment, we stress that most differential distributions are strongly similar in the two polarization 
definitions. However, there are kinematic observables,
like the charged lepton {pseudo}rapidity, for which the distributions are noticeably different.
We will return to this point in the next section.

\section{Results with lepton cuts}\label{leptoncut}
Imposing cuts on leptons is unavoidable at the LHC. Therefore, for our purposes, it is essential to
study the kinematic distributions for polarized intermediate $W$ bosons in the presence of transverse 
momentum and {pseudo}rapidity cuts on the final state leptons, to address the extraction
of polarization fractions and polarized cross-sections in a realistic setup.
In this section, the complete set of selection cuts described in Sect.~\ref{setup} is applied.

\begin{figure*}[hbt]
\centering
\includegraphics[scale=0.8]{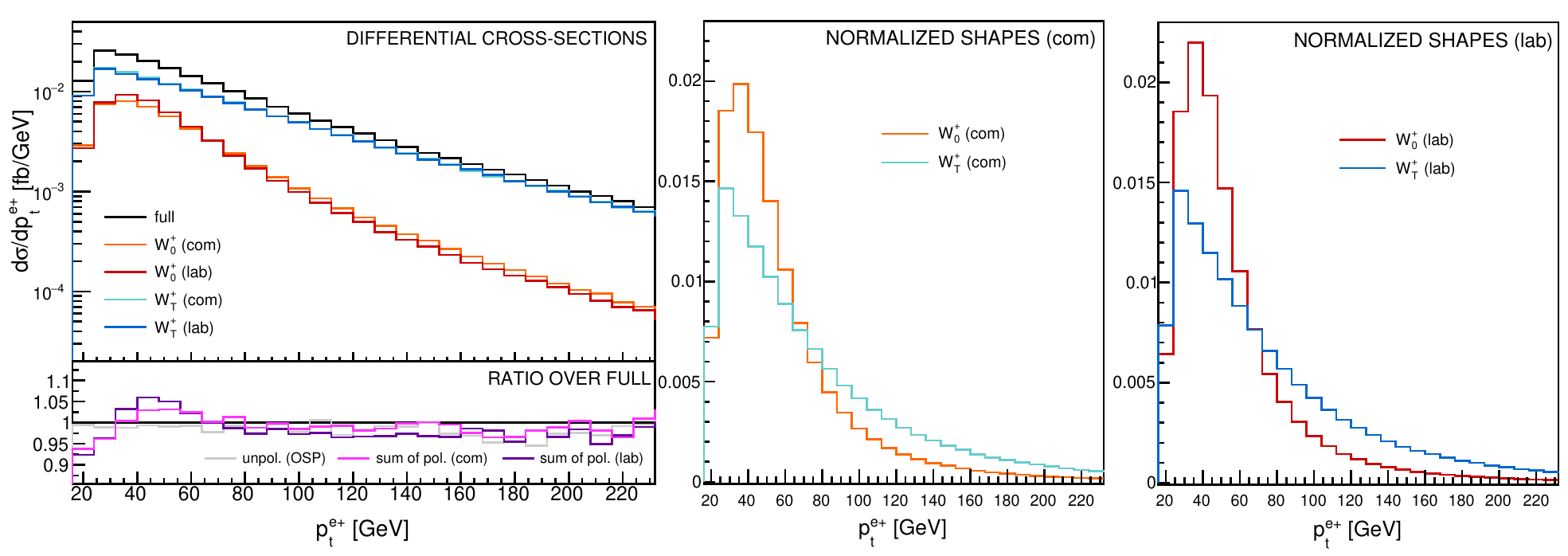}
\caption{Distributions in the positron transverse momentum, in the presence of lepton cuts.
The $W^+$ decaying into $e^+\nu_e$ has definite polarization state, while the one decaying into 
$\mu^+\nu_\mu$ is unpolarized.
The polarizations are defined in the CoM frame of the $WW$ system (com) or in the laboratory frame (lab).
The figure is organized as follows: differential distributions (top left), ratio over the full result (bottom left),
distribution shapes normalized to have unit integral for polarized signals defined in the \wwcom~(middle)
and in the Lab (right).}\label{pte}
\end{figure*}

\begin{figure*}[htb]
\centering
\includegraphics[scale=0.8]{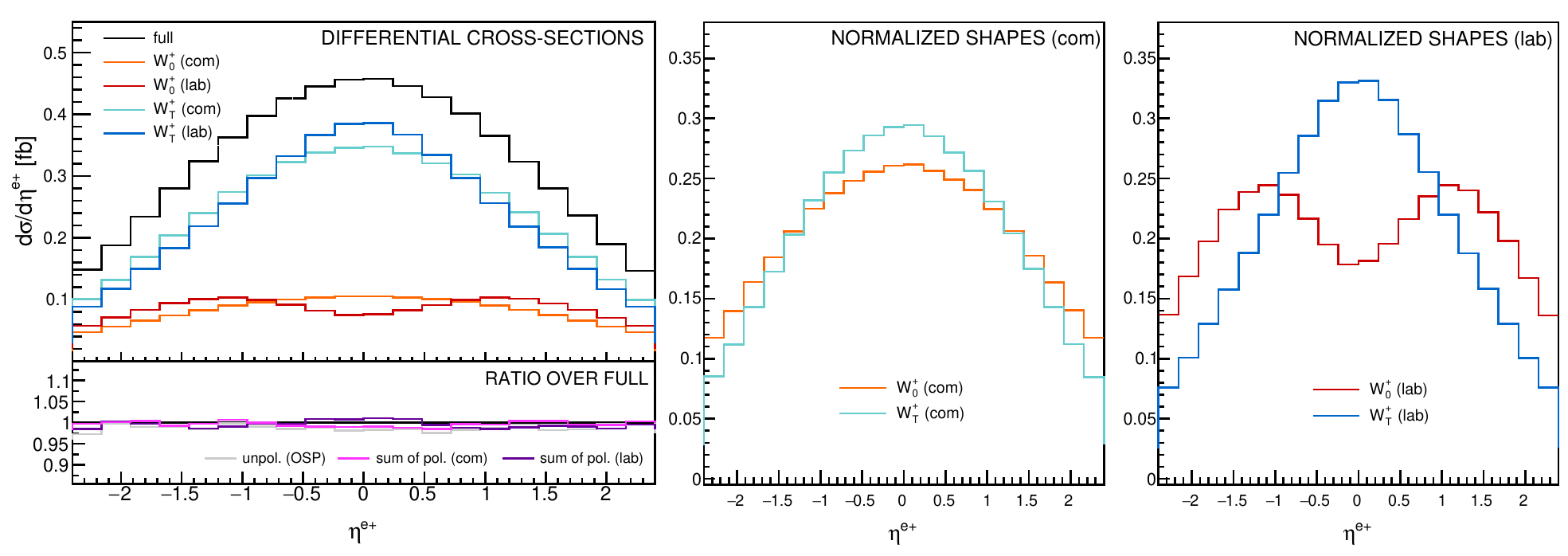}
\caption{Distributions in the positron pseudorapidity, in the presence of lepton cuts.
The figure is structured as Fig.~\ref{pte}.}\label{etae}
\end{figure*}

In analogy with Sect.~\ref{inclusive}, we start by presenting in Tab.~\ref{table:sigmalepmww161}
the total cross-sections for singly- and doubly-polarized VBS signals. 
\begin{table}[th]
\begin{center}
\begin{tabular}{|C{1.4cm}||C{1.65cm}|C{1.65cm}|C{1.2cm}|}
\hline
\cellcolor{ForestGreen!9}   & \cellcolor{ForestGreen!9}  Lab  & \cellcolor{ForestGreen!9}  \wwcom &  
\cellcolor{ForestGreen!9} ratio    \\
\hline
\hline
full   & \multicolumn{2}{c|}{1.593(2) } & -  \\
\hline                                  
unpol & \multicolumn{2}{c|}{1.572(2) } & -  \\
\hline                                  
\hline                                 
0-unpol &  0.4226(4)  &      0.4036(5)     & 0.96 \\
\hline                                  
T-unpol &  1.165(1)  &      1.182(2)     &  1.01 \\
\hline                                  
\hline                                  
0-0     &  0.1185(1)  &      0.1552(2)     & 1.31   \\
\hline                   
0-T,T-0    & 0.3062(3)   &      0.2519(3)     &  0.82  \\
\hline                   
T-T     &  0.8690(9)  &      0.9350(9)     & 1.08 \\
\hline
\end{tabular}
\end{center}
\caption{Total cross-sections (fb) or $W^+W^+$ scattering in the presence of lepton cuts. Same notation
as in
Tab.~\ref{table:sigmaincl}.}
\label{table:sigmalepmww161}
\end{table}
It is well known \cite{Stirling:2012zt,Belyaev:2013nla,Ballestrero:2017bxn}
that lepton cuts spoil the cancellation of
interference terms between different polarization modes.
In fact, the sum of singly- or doubly-polarized cross-sections is roughly 1.5\% larger than the
OSP unpolarized one, signaling small negative interferences. The off-shell effects that are missing
in the OSP approximation are of the same order of magnitude but positive. This results in a
sum of polarized signals which reproduces to less than 0.5\% the full cross-section.
Applying lepton cuts reduces all cross-sections by roughly a factor of two, but does not
sizeably change the polarization fractions obtained in the inclusive setup.
The differences between the two polarization definitions are essentially unchanged, apart from a mild
enhancement in the longitudinal-longitudinal contribution in the \wwcom~ definition (+31\%, w.r.t. the Lab).
The doubly-longitudinal fraction is 7.5\%(10\%), while the doubly-transverse
cross-section is the dominant one, as expected, accounting for the 55\%(59\%) of the total
in the Lab(\wwcom). Each of the two mixed contributions accounts for the remaining 19\%(16\%).

As already observed in the inclusive case, the singly-polarized signals are rather insensitive
to the polarization definition.
The singly-longitudinal(transverse) fraction accounts for the 26\%(70\%) of the total, in both definitions.

Since both are experimentally interesting,
in the following we present singly- and doubly-polarized differential distributions for some relevant 
variables.

Kinematic observables which depend on the decay products of a single $W$ are natural choices for the 
extraction of the polarized components of a single boson.

\begin{figure*}[htb]
\centering
\includegraphics[scale=0.8]{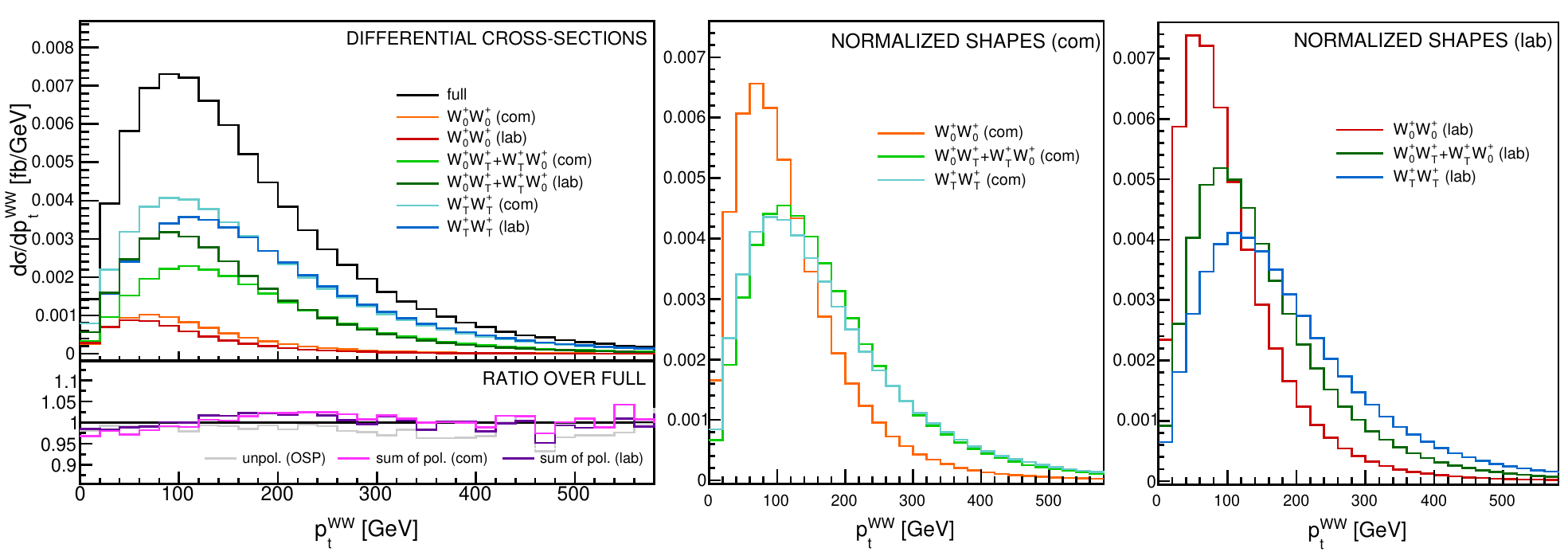}
\caption{Distributions in the four lepton system transverse momentum, in the presence of lepton cuts.
Both $W^+$ bosons have definite polarization state. The polarizations are defined in the CoM frame
of the $WW$ system (com) or in the laboratory frame (lab). The figure is organized as follows:
differential distributions (top left), ratio over the full result (bottom left), distribution
shapes normalized to unit integral for polarized signals defined in the \wwcom~(middle) and in the Lab (right).}\label{ptww}
\end{figure*}

In Fig.~\ref{pte} we consider the transverse momentum of the
positron.
These distributions confirm that the differences between the two polarization definitions are very small,
particularly for the transverse component, for which even the normalized shapes are almost identical. 
The longitudinal component shows a 4\% discrepancy at the integrated level.
The shapes are also different: the distribution in the Lab is slightly
narrower around the peak at 35 GeV.

The distribution of the positron transverse momentum for longitudinal and transverse polarization are 
significantly different.
The former has a large peak around 35 GeV, while the latter is harder, with a smaller peak close
to the threshold at 20 GeV.
 
The interferences are generally small, reaching 5-6\% in the soft region of the spectrum.
They change sign at about 30 GeV. The OSP description of the unpolarized process is good, even in the tail 
of the distribution.

The positron pseudorapidity is shown in Fig.~\ref{etae}. 
The OSP technique performs very well, and the interferences are below 2\% in the whole range 
$|\eta_\ell|<2.5$.
Although the integrated cross-section for singly-polarized configurations gives similar results in the
two definitions, there are relevant differences in the distributions.
In the \wwcom, the longitudinal and transverse
distributions feature a maximum at zero pseudorapidity.
The transverse shape is narrower.
On the contrary, in the Lab, the two polarization modes are quite distinct. The transverse distribution is 
similar to the \wwcom~one,  while the longitudinal has two peaks for
$|\eta_{e^+}|\approx 1$ and a local minimum at zero pseudorapidity. 
In this case it is clearly easier to separate the longitudinal from the transverse component if the polarizations
are defined in the Lab.
The positron pseudorapidity is the first evidence that different polarization definitions can give very different 
results for some kinematic variables, even if the polarization fractions are similar.
Therefore a detailed comparison of the full set of experimental observables with different polarization 
definitions should be performed.

We now move to
doubly-polarized distributions, in order to gain a more detailed understanding of the spin structure
of same sign off-shell $W$ bosons produced in VBS at the LHC.

 In the following we present the differential cross-sections for two observables that depend on the
charged lepton and neutrino kinematics, and that are symmetric under the exchange of the two lepton 
flavours.
This means that the longitudinal-transverse and transverse-longitudinal distributions are equal.
Therefore, we sum them in a single mixed polarization result.

\begin{figure*}[htb]
\centering
\includegraphics[scale=0.8]{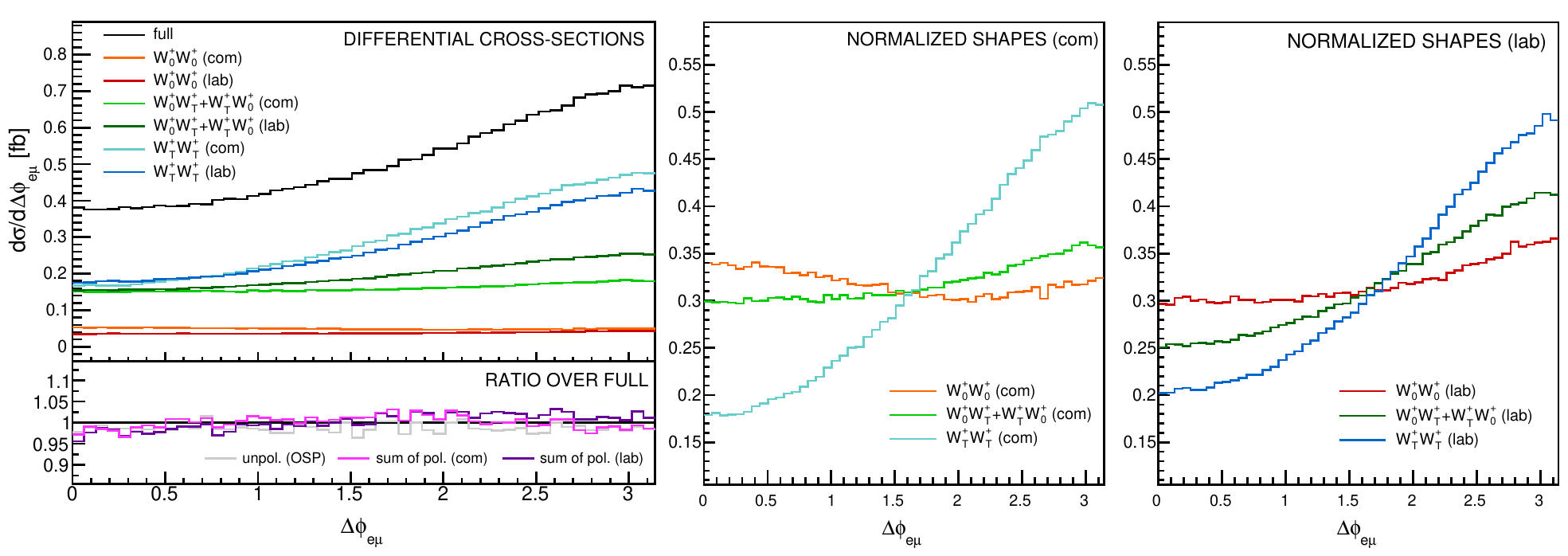}
\caption{Distributions in the azimuthal separation between the two charged leptons, 
in the presence of lepton cuts. The figure is structured as Fig.~\ref{ptww}.}\label{deltaphi}.
\end{figure*}
 
In Fig.~\ref{ptww} we consider the vector sum of the charged leptons transverse momenta and of the missing 
transverse momentum,  which is accessible at the LHC.
Since there is no explicit cut on this variable, it might be better suited for the study
of the shapes of distributions with the purpose of discriminating among polarization states.
This observable is very well described by the doubly resonant OSP unpolarized calculation,
as the discrepancies with respect to the full distribution amount at a few percent in all parts of the spectrum.
The interferences are almost negligible in the soft region, small, less than  5\%, and negative for
$p_t^{WW}>150\GeV$.
The transverse and mixed distributions in the \wwcom~differ considerably from
the ones in the Lab, mostly in the region around the peak.
In this range, the
noticeable enhancement of the doubly-transverse component in the \wwcom~is counter-balanced by 
the reduction of the mixed ones.
On the contrary, a sizeable enhancement of the doubly-longitudinal component in the \wwcom~arises
above the peak, for $p_t^{WW}> 60 \GeV$.
In the \wwcom, the distribution shapes for the doubly-transverse and mixed contributions are quite
similar, and feature a peak at $p_t^{WW}\approx 100 \GeV$. In the Lab, the mixed polarization
distribution is narrower than the doubly-transverse one.
In both polarization definitions, the  longitudinal-longitudinal distribution peaks at 
lower values than the other
components, for $p_t^{WW}\approx 70 (80)\GeV$ in the Lab (\wwcom).
 
It should be mentioned that, at LO, the four-lepton system transverse momentum coincides with the
transverse momentum of the two tagging jets. 
Therefore, this variable is expected to be sensitive to additional QCD radiation.

The marked shape differences among the doubly-polarized states, the small interferences and the good 
OSP description make this observable well suited for extracting polarizations, 
provided higher-order perturbative corrections, parton-shower and detector effects do not disrupt too much
the distribution shapes.

In the last part of this section we consider the azimuthal separation between the two charged leptons 
$\Delta\phi_{e\mu}$.
This angular variable is expected to be measured with high accuracy at the LHC. 
In Fig.~\ref{deltaphi} we show the doubly-polarized distributions.
Interferences among polarizations and non-resonant effects never exceed 4\%.
Even though interference terms are small, in our experience \cite{Ballestrero:2017bxn,Ballestrero:2019qoy}, it 
is useful to include them in fitting the data with polarized distributions.
Note that large interferences in $\Delta\phi$ can arise in other processes \cite{Denner:2020bcz,Maina:2020rgd}.

The unpolarized distributions and those involving at least one transverse boson, peak at
$\Delta\phi_{e\mu} = \pi$.
The doubly-longitudinal component is flatter in both polarization definitions. The mixed 
contribution is rather flat in the \wwcom, while in the Lab its shape is intermediate between the longitudinal 
and the transverse components.
All  the polarized distributions are approximately symmetric
about $\Delta\phi_{e\mu} = \pi/2$. 
However, the shapes are noticeably different for the three
doubly-polarized states, particularly in the \wwcom. Despite a lower discrimination power,
the sensitivity of $\Delta\phi_{e\mu}$ to the different polarization states is evident even in the Lab.
 
We have examined a number of other kinematic observables. They are not presented here
because they have less discriminating power than the variables we have shown above.
As an example, we have examined the $R_{p_t}$
variable proposed in Ref.~\cite{Doroba:2012pd} to improve the sensitivity to new physics effects.
We found that the shape of the $R_{p_t}$ distribution is roughly 
the same for longitudinal and transverse $W$ bosons.

We only investigated a limited number of observables, one at a time, at parton level. A more comprehensive
and realistic study is clearly warranted in order to optimize the multi-variate analysis of available data.

\section{A quick review of other VBS processes.}\label{otherproc}
Even though this work focuses on the scattering of two same sign $W$ bosons, it is interesting
to investigate whether the two different polarization definitions give different results also in the other 
vector boson scattering processes at the LHC.
All the results of Refs.~\cite{Ballestrero:2017bxn,Ballestrero:2019qoy} have been obtained with
polarization vectors defined in the Lab.
In this section we present polarized total cross-sections for the following processes:
 \begin{itemize}
 \item $p p \rightarrow j j \,e^+ {\nu_e} \mu^- \overline{\nu}_\mu \quad (W^+ W^-)$; \\[-0.6cm]
 \item $p p \rightarrow j j \,e^+ {\nu_e} \mu^+ \mu^- \quad (W^+ Z)$; \\[-0.6cm]
 \item $p p \rightarrow j j \,e^+ e^- \mu^+ \mu^- \quad (Z   Z)$.\\[-0.6cm]
 \end{itemize}

We have simulated $W^+W^-$ scattering using the inclusive setup described in Sect.~\ref{setup}. 
Since the OSP results
(polarized and unpolarized) are naturally restricted to the region $M_{WW}>2M_W$,  we have 
applied the same cut to the full calculation. This excludes the contribution of $s$-channel Higgs production 
and allows a meaningful comparison between the two results.
While a direct cut on the mass of the $WW$ system is unrealistic, the  Higgs signal can be excluded by 
appropriate  cuts on the leptonic system and the missing momentum.
The OSP calculation reproduces the full one very accurately.

 In Tab.~\ref{table:sigmawpwm} we show the polarized and unpolarized total cross-sections for opposite-sign 
 $W$ scattering.

 \begin{table}[ht!]
\begin{center}
\begin{tabular}{|C{1.4cm}||C{1.65cm}|C{1.65cm}|C{1.2cm}|}
\hline
\cellcolor{ForestGreen!9}   & \cellcolor{ForestGreen!9}  Lab  & \cellcolor{ForestGreen!9}  $WW$ CoM &  \cellcolor{ForestGreen!9} ratio    \\
\hline
\hline
full   & \multicolumn{2}{c|}{4.651(2) } & -  \\
\hline                                  
unpol & \multicolumn{2}{c|}{4.641(2) } & -  \\
\hline                                  
\hline                                 
0-unpol & 1.186(1) &  1.146(1)       & 0.97  \\
\hline                                  
T-unpol & 3.456(2) &  3.494(2)       & 1.01  \\
\hline                                 
unpol-0 & 1.2226(4)  &    1.1905(5)     & 0.97  \\
\hline                  
unpol-T & 3.418(1) &     3.450(1)    &  1.01 \\
\hline                                  
\hline                                  
0-0     & 0.3314(2)  & 0.3786(3)         & 1.14  \\
\hline                                  
0-T     & 0.8545(4)  & 0.7669(3)         & 0.90  \\
\hline                                  
T-0     & 0.8912(4)  & 0.8119(4)         & 0.91  \\
\hline                                 
T-T     & 2.563(1)   & 2.683(1)          & 1.05  \\
\hline
\end{tabular}
\end{center}
\caption{Total cross-sections (fb) for $W^+W^-$  scattering in the absence of lepton cuts, with 
$M_{4\ell}>2M_{W}\GeV$.
}
\label{table:sigmawpwm}
\end{table}

The singly-polarized cross-sections are very similar in the two definitions. They differ by about 1\% for the
transverse case and by  3\% for the longitudinal one. These results are in line with those obtained in 
same-sign $WW$ scattering, suggesting that such effects are not specific to $W^+W^+$.

Doubly-polarized signals also show a trend similar to the one for same-sign bosons.
The doubly-longitudinal \wwcom~signal is 14\% larger than the one in the Lab, while the purely transverse
is larger by 5\% . The mixed signals in the \wwcom~are 9\% smaller than in the Lab.

Encouraged by the strong similarity of the results obtained for both same-sign and opposite-sign $WW$ 
scattering, one expects the same effects to appear in other VBS channels.
For both $W^+Z$ and $ZZ$, we impose the jet cuts introduced in Sect.~\ref{setup},
and a minimum invariant mass cut of 200 GeV on the four lepton system.
The total cross-sections for
$ZZ$ and $W^+Z$ are shown in Tabs.~\ref{table:sigmaincl_ZZ} and \ref{table:sigmaincl_WZ}, respectively.
For $ZZ$ processes, the differences between the two definitions of polarizations are roughly the same as in 
$W^+W^+$, both for singly-polarized and doubly-polarized total cross-sections, as can be observed 
comparing Tab.~\ref{table:sigmaincl_ZZ} with Tab.~\ref{table:sigmaincl}.
\begin{table}[ht!]
\begin{center}
\begin{tabular}{|C{1.4cm}||C{1.65cm}|C{1.65cm}|C{1.2cm}|}
\hline
\cellcolor{ForestGreen!9}   & \cellcolor{ForestGreen!9}  Lab  & \cellcolor{ForestGreen!9}  $ZZ$ CoM &  
\cellcolor{ForestGreen!9} ratio    \\
\hline
\hline
full   & \multicolumn{2}{c|}{0.1270(1) } & -  \\
\hline                                  
unpol & \multicolumn{2}{c|}{0.1264(1) } & -  \\
\hline                                  
\hline                                 
0-unpol & 0.03328(2) & 0.03104(2)        & 0.93  \\
\hline                                  
T-unpol & 0.09295(5) & 0.09511(4)         & 1.02   \\
\hline                                  
\hline                                  
0-0     & 0.00910(1) & 0.01087(2)         &  1.19 \\
\hline                                  
0-T,T-0     & 0.02421(2) & 0.02022(2)          &  0.84 \\
\hline                                 
T-T     & 0.06869(3) & 0.07490(4)         &  1.09 \\
\hline
\end{tabular}
\end{center}
\caption{Total cross-sections (fb) for $ZZ$ scattering in the absence of lepton cuts, with 
$M_{4\ell}>200\GeV$.
}
\label{table:sigmaincl_ZZ}
\end{table}

\begin{table}[ht!]
\begin{center}
\begin{tabular}{|C{1.4cm}||C{1.65cm}|C{1.65cm}|C{1.2cm}|}
\hline
\cellcolor{ForestGreen!9}   & \cellcolor{ForestGreen!9}  Lab  & \cellcolor{ForestGreen!9}  $WZ$ CoM &  
\cellcolor{ForestGreen!9} ratio    \\
\hline
\hline
full   & \multicolumn{2}{c|}{ 0.5253(3)} & -  \\
\hline                                  
unpol & \multicolumn{2}{c|}{ 0.5210(3)} & -  \\
\hline                                  
\hline                                 
0-unpol & 0.1216(1)  & 0.1292(1)        &  1.06 \\
\hline                                  
T-unpol & 0.3992(2) & 0.3918(3)         &  0.98 \\
\hline                                 
unpol-0 & 0.1370(1)  &  0.1436(1)       &  1.05 \\
\hline                                  
unpol-T & 0.3839(2) &  0.3773(2)        &  0.98 \\
\hline                                  
\hline                                  
0-0     & 0.03236(3)  & 0.03993(5)         & 1.23  \\
\hline                                  
0-T     & 0.08923(8) & 0.08926(8)          & 1.00  \\
\hline                                  
T-0     & 0.1045(1) & 0.1039(1)         &  0.99 \\
\hline                                 
T-T     & 0.2948(2) & 0.2876(2)         &  0.98 \\
\hline
\end{tabular}
\end{center}
\caption{Total cross-sections (fb) for $W^+Z$ scattering in the absence of lepton cuts, 
with $M_{4\ell}>200\GeV$.
}
\label{table:sigmaincl_WZ}
\end{table}

A detailed analysis of Tab.~\ref{table:sigmaincl_WZ}  reveals new aspects that characterize 
polarized signals in the $W^+Z$ process.
The singly-longitudinal cross-section in the \wzcom~is 5-6\%
larger than in the Lab, both for the $W^+$ and for the $Z$ boson. This is
balanced by the opposite behavior of the dominant transverse contribution, which is 
2\% smaller in the \wzcom. These differences between the \wzcom~and the Lab,
though not large,  go in the opposite direction with respect to the $WW$ and $ZZ$ processes.

Furthermore, the doubly-longitudinal signal in the \wzcom~is about 20\% larger than in the Lab.
This enhancement is balanced by the decrease of the transverse-transverse contribution,
and not of the mixed
ones as in $WW$ processes. The mixed configurations have almost exactly the
same cross-section with both definitions ($\lesssim 0.5\%$ differences).

As a last result of this section, we have found that the differences
between the two polarizations definitions in same-sign $WW$ are not peculiar to the
SM dynamics, but are present even in the strongly interacting Higgsless model
(SM with $M_H\rightarrow \infty$).  
Using the same cuts as in Sect.~\ref{inclusive},
we have checked numerically that in the Higgsless model,
the longitudinal-longitudinal signal is much larger than the SM one, both in the \wwcom~and in 
the Lab (+80\%).
Such a large deviation is motivated by the unitarity violations that characterize the high-energy regime,
but show up even imposing the loosest possible cut on the $WW$ mass ($M_{WW}>161\GeV$),  which is
needed for the OSP to provide a good description of the full results.
 
\section{Conclusions}\label{concl}
In this paper, we have investigated the phenomenology of polarized
vector bosons in $W^+W^+$ scattering. We have compared the results obtained defining
the electroweak boson polarization vectors in the \wwcom~and in the Lab reference frames.

The complex structure of VBS processes makes it difficult to explain analytically
the differences between the results in the  \wwcom~and the Lab in terms of resonant 
matrix-elements and  Lorentz transformations.
Therefore, only numerical studies, like the one presented
in this paper, can provide a detailed picture of the polarization structure of multi-boson processes, and 
identify a set of variables which permits, in a given choice of polarization vectors,
to separate the vector boson polarization states in the experimental analyses of  LHC data.

We have presented total and differential
cross-sections for both singly-polarized and doubly-polarized signals.
The singly-polarized signals are weakly sensitive to the polarization definition at the
integrated level. The transverse component is typically three times larger than the longitudinal one.
The distributions for some kinematic 
variables, like the single lepton transverse momentum and pseudorapidity, depend noticeably on the $W$ 
polarization and, therefore, can be used to separate the different contributions and
determine the polarization fractions.

The longitudinal-longitudinal and transverse-transverse
cross-sections are enhanced by defining polarizations in the \wwcom, while mixed configurations are 
diminished. This might be useful for searches of deviations from the SM description of EWSB.
The differences between the distributions for transverse and longitudinal polarizations in the Lab
are slightly larger than in the \wwcom.

Our results do not clearly favor either of the reference frames we have examined for the definition
of polarization vectors.

Beyond the specific study of same-sign $WW$ scattering, we have also compared the polarized total
cross-sections in the two definitions for other VBS channels. We have found strong similarities with 
$W^+W^+$ in  $W^+W^-$ and  $ZZ$ processes, while new features appear in $W^+Z$.

\section*{Acknowledgements}
We thank Ansgar Denner and Lucia Di Ciaccio for useful discussions.
The authors acknowledge the VBSCan COST Action CA16108.
EM is supported by the SPIF INFN project (Precision Studies of Fundamental Interactions),
GP is supported by the German Federal Ministry for Education and Research (BMBF) under contract 
no.~05H18WWCA1.

 \bibliographystyle{elsarticle-num} 
 \biboptions{numbers,sort&compress}
 \bibliography{pol3}

\end{document}